\documentclass[prd,aps,preprintnumbers,nofootinbib,superscriptaddress]{revtex4}
\pdfoutput=1
\usepackage{amssymb,amsmath,graphicx,bm}
\usepackage[final]{hyperref}

\usepackage[usenames,dvipsnames]{color}
\usepackage[normalem]{ulem} % \sout{old text} for strikeout
\renewcommand\sout{\bgroup \color[rgb]{0.55,0.00,0.99} \ULdepth=-.5ex \ULset}

\newcommand{\simorder}{\raisebox{-4pt}{$\, \stackrel{\textstyle <}{\sim} \,$}}

\begin{document}

\title{GTMD model predictions for diffractive dijet production at EIC}

\author{Dani\"el Boer}
\email{d.boer@rug.nl}
\affiliation{ {Van Swinderen Institute for Particle Physics and Gravity, University of Groningen, Nijenborgh 4, 9747 AG Groningen, The Netherlands}}

\author{Chalis Setyadi}
\email{c.setyadi@rug.nl; chalis@ugm.ac.id}
\affiliation{ {Van Swinderen Institute for Particle Physics and Gravity, University of Groningen, Nijenborgh 4, 9747 AG Groningen, The Netherlands}}
\affiliation{ {Department of Physics, Universitas Gadjah Mada, BLS 21 Yogyakarta, Indonesia
}}

\begin{abstract}
In this paper we consider a small-$x$ model for gluon GTMDs that we fit to data on diffractive dijet production in electron-proton collisions obtained by HERA's H1 Collaboration. Assuming a small number of free parameters, each with a physical motivation, we are able to describe those data fairly well and with this model we obtain predictions for the EIC for both electroproduction and photoproduction which may allow to further test the underlying GTMD description. In the general discussion of the impact parameter dependence we recall some subtle issues related to localization of states, choice of frames, and discuss what these aspects imply for the range of applicability of the model.
\end{abstract}

%\pacs{12.38.-t; 13.85.Ni; 13.88.+e}
\date{\today}

\maketitle

\section{Introduction}
Generalized Transverse Momentum Dependent parton distributions (GTMDs) and the associated Wigner parton distributions were first considered in \cite{Ji:2003ak,Belitsky:2003nz} and analyzed further in e.g.\ \cite{Meissner:2008ay,Meissner:2009ww,Lorce:2011dv,Echevarria:2016mrc} for quarks and in \cite{Echevarria:2016mrc,More:2017zqp} for gluons. The first suggestion to access GTMDs experimentally was put forward in \cite{Hatta:2016dxp}. In that paper the process of diffractive dijet production in electron-proton collisions was considered to probe gluon GTMDs. Diffractive dijet production was earlier suggested as a way to probe gluon Generalized Parton Distributions (GPDs) \cite{Braun:2005rg} and considered at small $x$ in \cite{Altinoluk:2015dpi}. Diffractive single jet production was studied in \cite{Peccini:2020tpj}. In the present paper we follow up on these ideas and construct a model for the unpolarized gluon GTMD similar to the one put forward in \cite{Hatta:2016dxp}. Models for quark GTMDs have been considered in e.g.\ \cite{Kaur:2019kpi,Luo:2020yqj,Zhang:2020ecj}. For gluon GTMDs the models are so-far based on the small-$x$
McLerran-Venugopalan (MV) model \cite{McLerran:1993ni,McLerran:1993ka,McLerran:1994vd} and related Color Glass Condensate (CGC) descriptions. We will also take the MV model as our starting point, but introduce a few free parameters to be fitted to H1 data from the HERA collider in order to arrive at predictions for the U.S.-based Electron-Ion Collider (EIC) 
that hopefully will allow to further test the underlying GTMD description.
We will only consider unpolarized gluons, because the azimuthal modulations in the diffractive dijet cross section arising from the elliptic GTMD \cite{Hatta:2016dxp,Zhou:2016rnt} are expected to be much smaller than the present cross section uncertainties. In the model studies of \cite{Hagiwara:2016kam,Hagiwara:2017fye,Mantysaari:2019csc,Salazar:2019ncp,Hatta:2020bgy} and in the CMS data \cite{CMS:2020ekd} the azimuthal asymmetries are found to be at the 10-30\% level or (much) smaller. As a first step it would be important to check whether the GTMD description of diffractive dijet production is consistent with cross section measurements in various kinematic variables and various kinematic regions. With the presented results we hope to facilitate such a study.

This paper is organized as follows. In section \ref{GTMDsintro} we discuss the definitions and properties of GTMDs in general and recall some subtle issues regarding the localization of states and the impact parameter dependence\footnote{We thank Markus Diehl for sharing his insights concerning these aspects.}. In section \ref{model} we discuss the model and its free parameters, addressing also the range of applicability of the model. We furthermore argue that the process of diffractive dijet production in electron-proton collisions in the so-called correlation limit falls within that range.
In section \ref{crosssection} we provide details on the cross section calculation in terms of the unpolarized gluon GTMD. We discuss the fit of our model to H1 data in section \ref{fits} and present our predictions for the EIC in section \ref{predictions}. We end with a concluding section. 

\section{Definitions of GTMDs\label{GTMDsintro}}
GTMDs are 5-dimensional parton distributions that depend on the parton's lightcone momentum fraction $x$, the parton's transverse momentum $\bm{k}_\perp$ and the transverse off-forwardness $\Delta_\perp$ by which the incoming hadron momentum gets modified. The associated Wigner distribution parton is a function of $x$,  $\bm{k}_\perp$, and the impact parameter $\bm{b}_\perp$ which is the Fourier conjugate of $\bm{\Delta}_\perp$. 
GTMDs can be viewed as off-forward extensions of Transverse Momentum Dependent parton distributions (TMDs) or as transverse momentum dependent extensions of GPDs. As a consequence, the GTMDs inherit properties of both TMDs and GPDs and any subtle issues related to them.  
In this section we will go into some of these matters, restricting the discussion to the distribution of unpolarized quarks inside an unpolarized hadron, for which we take a proton for definiteness. 

First of all, the quark GTMD $q(x,\bm{k}_\perp,\bm{\Delta}_\perp)$ can be defined as the off-forward generalization of the quark TMD $q(x,\bm{k}_\perp)$:
\begin{align}
q(x,\bm{k}_\perp,\bm{\Delta}_\perp) & \equiv  \int \frac{d \lambda}{2\pi P^+}\, d^2 \bm{r}_\perp \, e^{i \lambda x} e^{i\bm{k}_\perp\cdot\bm{r}_\perp}\, \langle {P'} \vert \;  {\overline \psi(-\frac{\lambda}{2}n-\frac{\bm{r}_\perp}{2}) \, \gamma^+ {\cal L}
\, \psi(\frac{\lambda}{2}n+\frac{\bm{r}_\perp}{2})} \; \vert {P} \rangle , 
\label{GTMDdeffromTMD}
\end{align}
where the lightlike vector $n$ specifies the $-$ direction, whereas the proton momentum $P$ specifies the $+$ direction: $P\cdot n=P^+$. In the above expression $\Delta= P'-P$ denotes the off-forwardness considered here for zero skewness, i.e.\ $\xi = - \Delta^+/({P'}^+ + P^+)=0$ and $\Delta= \Delta_\perp$. The path-ordered exponential or gauge link ${\cal L}$ does not play an important role here and will be left unspecified. 

Alternatively, the quark GTMD can be defined as the Fourier transform of the Wigner quark distribution $W(x,\bm{k}_\perp, \bm{b}_\perp)$, which itself can be defined as the transverse momentum dependent generalization of the impact parameter dependent GPD $q(x, \bm{b}_\perp)$ \cite{Soper:1976jc,Burkardt:2000za,Diehl:2002he,Burkardt:2002hr,Burkardt:2002ks}: 
\begin{align}
q(x, \bm{b}_\perp) = \int \frac{d \lambda}{2\pi P^+}\, e^{i \lambda x}\, 
\langle {P^+, \bm{R}_\perp = 0} \vert\;  
\overline \psi(-\frac{\lambda}{2}n+{\bm{b}_\perp}) \, \gamma^+ \, {\cal L}
\, \psi(\frac{\lambda}{2}n+{\bm{b}_\perp}) \; 
\vert {P^+, \bm{R}_\perp = 0} \rangle, 
\label{qxb}
\end{align}
where the impact parameter $\bm{b}_\perp$ is measured w.r.t.\ the transverse center of longitudinal momentum 
$\bm{R}_\perp^{CM} \equiv \sum_i x_i \bm{r}_{\perp i}$ of the system and $\vert {P^+, \bm{R}_\perp = 0} \rangle$ is the normalized proton state localized 
in the spatial $\perp$ direction \cite{Diehl:2002he,Burkardt:2002ks}:
\begin{align}
\vert {P^+, \bm{R}_\perp = 0} \rangle = {\cal N} \int \frac{d^2 \bm{P}_\perp}{(2\pi)^2} \Phi(\bm{P}_\perp) |P^+,\bm{P}_\perp \rangle,
\end{align}
for some wave packet $\Phi(\bm{P}_\perp)$. This expression for $q(x, \bm{b}_\perp)$ thus depends on the wave packet considered. 
If this wave packet is sufficiently localized in transverse position space, such that $\Phi(\bm{P}_\perp+\bm{\Delta}_\perp) \approx \Phi(\bm{P}_\perp)$, meaning it is slowly varying on the scale of the off-forwardness, one can relate it to the standard GPD $H$ \cite{Burkardt:2000za,Diehl:2002he,Burkardt:2002hr,Burkardt:2002ks}:
\begin{align}
q(x, \bm{b}_\perp) & = |{\cal N}|^2 \int \frac{d \lambda}{2\pi P^+}\, e^{i \lambda x}\int \frac{d^2\bm{P}_\perp}{(2\pi)^2} \frac{d^2\bm{P}^\prime_\perp}{(2\pi)^2} \, \Phi^*(\bm{P}^\prime_\perp) \Phi(\bm{P}_\perp)   
\langle {P^+, \bm{P}^\prime_\perp} \vert\;  
\overline \psi(-\frac{\lambda}{2}n+{\bm{b}_\perp}) \, \gamma^+ \, {\cal L}
\, \psi(\frac{\lambda}{2}n+{\bm{b}_\perp}) \; 
\vert {P^+, \bm{P}_\perp} \rangle\nonumber\\
& = |{\cal N}|^2 \int \frac{d \lambda}{2\pi P^+}\, e^{i \lambda x}\int \frac{d^2\bm{P}_\perp}{(2\pi)^2} \frac{d^2\bm{P}^\prime_\perp}{(2\pi)^2} \, \Phi^*(\bm{P}^\prime_\perp) \Phi(\bm{P}_\perp)  e^{-i\bm{b}_\perp \cdot \bm{\Delta}_\perp} \langle {P^+, \bm{P}^\prime_\perp} \vert\;  
\overline \psi(-\frac{\lambda}{2}n) \, \gamma^+ \, {\cal L}
\, \psi(\frac{\lambda}{2}n) \; 
\vert {P^+, \bm{P}_\perp} \rangle\nonumber\\
& \approx |{\cal N}|^2 \int \frac{d^2\bm{P}_\perp}{(2\pi)^2} |\Phi(\bm{P}_\perp)|^2 \frac{d^2\bm{\Delta}_\perp}{(2\pi)^2} \, e^{-i\bm{b}_\perp \cdot \bm{\Delta}_\perp} H(x,0,-\bm{\Delta}^2_\perp) = \int \frac{d^2\bm{\Delta}_\perp}{(2\pi)^2} \, e^{-i\bm{b}_\perp \cdot \bm{\Delta}_\perp} H(x,0,-\bm{\Delta}^2_\perp).
\end{align}
In the second step we used that, unlike forward matrix elements, off-forward matrix elements of the form $\langle {P^+, \bm{P}^\prime_\perp} \vert\; O(\bm{b}_\perp) \; \vert {P^+, \bm{P}_\perp}\rangle$ are not translation invariant, but pick up a phase when translating the operator $O(\bm{b}_\perp)$ to $O(\bm{0}_\perp)$. In the above derivation it was also used that $
\langle P^+, \bm{P}^\prime_\perp \vert\;  O(\bm{0}_\perp) \;  \vert P^+, \bm{P}_\perp \rangle$ only depends on the difference of $\bm{P}^\prime_\perp$ and $\bm{P}_\perp$, which is a consequence of invariance under transverse boosts, see \cite{Diehl:2002he}, in particular its Eq.\ (5).

One observes that only in the case that $\Phi(\bm{P}_\perp)$ is a constant, the relation between $q(x,\bm{b}_\perp)$ and $H(x,0,-\bm{\Delta}^2_\perp)$ is exact. When viewing $q(x,\bm{b}_\perp)$ as the Fourier transform of the GPD $H(x,0,-\bm{\Delta}^2_\perp)$ it is thus understood that one considers a wave packet that is sufficiently localized in coordinate space and hence sufficiently delocalized in momentum space. This then raises the question of how to reconcile such a very delocalized state in transverse momentum space with a state that has a specific $z$-momentum and energy, which are related by $P^- = (M^2 + P_\perp^2)/(2P^+)$. This issue is known to pose a problem for 3D spatial distributions, where a state cannot be simultaneously in a definite eigenstate of position and momentum and frame and wave packet dependence enters\footnote{This issue received renewed attention recently in the context of defining the 3D charge radius for the nucleon \cite{Miller:2018ybm,Jaffe:2020ebz}. For nucleons (as opposed to heavy nuclei) the system size is not sufficiently large w.r.t.\ the Compton wavelength to allow for an unambiguous, wave packet independent, definition of the charge distribution and hence of the charge radius \cite{Jaffe:2020ebz}.}. For the 2D charge distribution and analogously for $q(x,\bm{b}_\perp)$ one can avoid this issue by boosting to a frame in which $P^+$ is much larger than the typical $P_\perp$ values. This allows to maintain $P^- = (M^2 + P_\perp^2)/(2P^+) \ll P^+$ in the wave packet, such that the state has sufficiently definite $P^0$ and $P^3$ components even if the $P_\perp$ distribution is very broad, cf.\ \cite{Diehl:2002he} for further discussion.
 
Starting from the impact parameter dependent GPD $q(x, \bm{b}_\perp)$ one obtains a definition of the Wigner parton distribution $W(x,\bm{k}_\perp, \bm{b}_\perp)$ defined as\footnote{In \cite{Ji:2003ak,Belitsky:2003nz} the Wigner quark distribution was defined as a generalization of the 3D charge density in the Breit frame (for a brief discussion of that see \cite{Boer:2004ai}), which inherits the mentioned wave packet dependence issue for the nucleon.}
\begin{align}
W(x, \bm{k}_\perp, \bm{b}_\perp) \equiv \int \frac{d \lambda}{2\pi P^+}\, d^2 \bm{r}_\perp \, e^{i \lambda x} e^{i\bm{k}_\perp\cdot\bm{r}_\perp}\, 
\langle {P^+, \bm{R}_\perp = 0} \vert\;  
\overline \psi(-\frac{\lambda}{2}n+{\bm{b}_\perp}-\frac{\bm{r}_\perp}{2}) \, \gamma^+ \, {\cal L}
\, \psi(\frac{\lambda}{2}n+{\bm{b}_\perp}+\frac{\bm{r}_\perp}{2}) \; 
\vert {P^+, \bm{R}_\perp = 0} \rangle. 
\end{align}
The GTMD is then defined as  
\begin{align}
q_W(x,\bm{k}_\perp,\bm{\Delta}_\perp) & \equiv \int \frac{d^2\bm{b}_\perp}{(2\pi)^2} \, e^{i\bm{b}_\perp \cdot \bm{\Delta}_\perp} 
W(x, \bm{k}_\perp, \bm{b}_\perp).
\label{GTMDdeffromWigner}
\end{align}
Just like for the Fourier transform of $q(x,\bm{b}_\perp)$ and the GPD $H(x,0,-\bm{\Delta}^2_\perp)$, one can equate the two GTMD definitions Eqs.\ (\ref{GTMDdeffromTMD}) and (\ref{GTMDdeffromWigner}) for a sufficiently narrow state $\vert {P^+, \bm{R}_\perp = 0} \rangle$ in coordinate space, 
for which, as we discussed above, one has to consider a frame in which $P^+$ is much larger than the typical $P_\perp$ values. We emphasize that the sufficiently narrow state refers to the wave packet in which the state is prepared, not to the nucleon or nucleus which itself will have some profile in transverse coordinate space that may be considerably less narrow.

When one considers small $x$ values gluons dominate, hence we will from now focus on (unpolarized) gluon distributions. More specifically, the process of diffractive dijet production in electron-proton or electron-nucleus collisions probes the dipole gluon GTMD (again considered for zero skewness) defined as \cite{Hatta:2016dxp,Boer:2018vdi}:
\begin{align}
G^{[+,-]}(x, \bm{k}_\perp, \bm{\Delta}_\perp) = \frac{2}{P^+} \int \frac{dz^- }{2\pi} \frac{d^2 \bm{z}_\perp}{(2\pi)^2} \, e^{ik\cdot z} \left.
\langle {P^\prime} \vert\; {\rm Tr}\left[F^{+i}\left(-\frac{z}{2}\right) U^{[+]} F^{+i}\left(\frac{z}{2}\right) U^{[-]} \right] \; \vert {P} \rangle \right|_{z^+=0}.
\end{align}
Here $i$ is a transverse index that is summed over in this case of unpolarized gluons and $U^{[\pm]}$ are the standard staple-like gauge links in the forward ($+$) and backward ($-$) lightcone directions, also simply referred to as $+$ and $-$ links.  The function $xG_{{\rm DP}}(x, \bm{q}_\perp, \bm{\Delta}_\perp)$ in \cite{Hatta:2016dxp} corresponds to $G^{[+,-]}(x, \bm{q}_\perp, \bm{\Delta}_\perp)$.

In the limit of $x \to 0$ one can show to arrive at \cite{Boer:2018vdi}
\begin{align}
G^{[+,-]}(\bm{k}_\perp, \bm{\Delta}_\perp) & = \frac{4}{g^2 \langle P | P\rangle} \int \frac{d^2 \bm{x}_\perp d^2 \bm{y}_\perp}{(2\pi)^3} \, 
e^{-i\bm{k}_\perp \cdot (\bm{x}_\perp-\bm{y}_\perp)} e^{i\bm{\Delta}_\perp \cdot (\bm{x}_\perp+\bm{y}_\perp)/2} 
\langle {P^\prime} \vert\; \partial^i_x \partial^i_y {\rm Tr}\left[U^{[\Box]}(\bm{y}_\perp,\bm{x}_\perp) \right] \; \vert {P} \rangle 
\nonumber \\
&\equiv \frac{1}{2\pi g^2} \left[\bm{k}_\perp^2-\frac{\bm{\Delta}_\perp^2}{4} \right] G^{[\Box]}(\bm{k}_\perp, \bm{\Delta}_\perp), 
\label{fromxytorb}
\end{align}
where $U^{[\Box]}(\bm{y}_\perp,\bm{x}_\perp)=U^{[+]}(\bm{y}_\perp,\bm{x}_\perp) U^{[-]}(\bm{x}_\perp,\bm{y}_\perp)$ will be referred to as the Wilson loop and $\langle P' | P\rangle = (2\pi)^3 2P^+ \delta(\Delta^+)\delta^{(2)}(\Delta_\perp)$ yielding the divergent factor $\langle P | P\rangle= 2P^+ \int db^- d^2 \bm{b}_\perp$, which is assumed to be regularized, e.g.\  by considering a finite volume. In this way one finds that
\begin{align}
G^{[\Box]}(\bm{k}_\perp, \bm{\Delta}_\perp) & = \frac{4N_c}{\langle P | P\rangle} \int \frac{d^2 \bm{x}_\perp d^2 \bm{y}_\perp}{(2\pi)^2} \, 
e^{-ik\cdot (\bm{x}_\perp-\bm{y}_\perp)} e^{i\Delta \cdot (\bm{x}_\perp+\bm{y}_\perp)/2} 
\langle {P^\prime} \vert\; S^{[\Box]}(\bm{x}_\perp,\bm{y}_\perp)  \; \vert {P} \rangle \nonumber\\
& =  \frac{4N_c}{\langle P | P\rangle} \int \frac{d^2 \bm{r}_\perp d^2 \bm{b}_\perp}{(2\pi)^2} \, 
e^{-i\bm{k}_\perp\cdot \bm{r}_\perp} e^{i\bm{\Delta}_\perp \cdot \bm{b}_\perp} 
\langle {P^\prime} \vert\; S^{[\Box]}(\bm{b}_\perp-\frac{\bm{r}_\perp}{2},\bm{b}_\perp+\frac{\bm{r}_\perp}{2})  \; \vert {P} \rangle 
\label{Gboxb}
\end{align}
where $S^{[\Box]}(\bm{x}_\perp,\bm{y}_\perp) \equiv {\rm Tr}\left[U^{[\Box]}(\bm{y}_\perp,\bm{x}_\perp)\right]/N_c$, $\bm{r}_\perp=\bm{y}_\perp-\bm{x}_\perp$ and $\bm{b}_\perp=(\bm{x}_\perp+\bm{y}_\perp)/2$. Comparing again to the notation of \cite{Hatta:2016dxp} we see that ${\cal F}_x = G^{[\Box]}/((4\pi)^2 N_c)$, where ${\cal F}_x$ satisfies the normalization condition $\int d^2 \bm{k}_\perp\,d^2 \bm{\Delta}_\perp \, e^{-i\bm{\Delta}_\perp \cdot \bm{b}_\perp}\,{\cal F}_x(\bm{k}_\perp, \bm{\Delta}_\perp)=1$.
In Eq.\ (\ref{Gboxb}) $\bm{b}_\perp$ is defined w.r.t.\ some unspecified reference point, so one may wonder what determines this position? In fact, in the derivation of Eq.\ (\ref{fromxytorb}) the following step is performed:
\begin{align}
\langle P+\Delta_\perp \vert O(\bm{0}_\perp,\bm{r}_\perp) \vert P \rangle = e^{i\bm{b}_\perp \cdot \bm{\Delta}_\perp} \langle P+\Delta_\perp \vert O(\bm{b}_\perp,\bm{r}_\perp) \vert P \rangle = \frac{\int d^2 \bm{b}_\perp e^{i\bm{b}_\perp \cdot \bm{\Delta}_\perp} \langle P+ \Delta_\perp \vert O(\bm{b}_\perp,\bm{r}_\perp) \vert P \rangle}{\int d^2\bm{b}_\perp} ,
\label{dbformal}
\end{align} 
where the last step is formally exact, but as mentioned, the normalization factor $\int d^2\bm{b}_\perp$ (which is part of $\langle P | P\rangle$) is actually divergent and requires consideration of a regulator. In the derivation it is used that although matrix elements of the form $\langle P+\Delta_\perp \vert O(\bm{b}_\perp) \vert P \rangle$ are both $\bm{b}_\perp$ and $\bm{\Delta}_\perp$ dependent, despite $\bm{b}_\perp$ and $\bm{\Delta}_\perp$ being each other's Fourier conjugates, the $\bm{b}_\perp$ dependence enters just through a phase. As a result, the integrand is actually $\bm{b}_\perp$ independent and any reference point will do.
However, the previous GPD analysis suggests that it is better to replace Eq.\ (\ref{dbformal}) by
\begin{align}
\langle P+\Delta_\perp \vert O(\bm{0}_\perp,\bm{r}_\perp) \vert P \rangle = \int \frac{d^2\bm{b}_\perp}{(2\pi)^2} \, e^{i\bm{b}_\perp \cdot \bm{\Delta}_\perp} \langle P^+, \bm{R}_\perp = 0 \vert  O(\bm{b}_\perp,\bm{r}_\perp) \vert P^+, \bm{R}_\perp = 0 \rangle,
\label{Rperpexpression}
\end{align}
where $\bm{b}_\perp$ is considered w.r.t.\ $\bm{R}_\perp=0$ and a large $P^+$ momentum frame and a spatially localized wave packet are implicitly considered. It is also implicitly used that $\langle {P^+, \bm{P}^\prime_\perp} \vert\;  O(\bm{0}_\perp,\bm{r}_\perp) \;  \vert {P^+, \bm{P}_\perp} \rangle$ only depends on the difference of $\bm{P}^\prime_\perp$ and $\bm{P}_\perp$, just like for $\bm{r}_\perp=0$ in the GPD case ($\bm{r}_\perp$ is not affected by the required transverse boosts, since the corresponding $r^+=0$). It is furthermore interesting to note that Eq.\ (\ref{Rperpexpression}) relates an off-forward matrix element to an integral over diagonal matrix elements. 

Following the above replacement, matrix elements of the 
form $\langle P^\prime \vert O(\bm{b}_\perp,\bm{r}_\perp)  \vert {P} \rangle/\langle P | P\rangle$ are to be interpreted as 
$\langle P^+, \bm{R}_\perp = 0 \vert  O(\bm{b}_\perp,\bm{r}_\perp) \vert P^+, \bm{R}_\perp = 0 \rangle$ 
in the expression for $G^{[\Box]}$ and, similarly, for ${\cal F}^{[\Box]}$ which follows from $G^{[\Box]}$ by the replacement $S^{[\Box]}(\bm{x}_\perp,\bm{y}_\perp) \to 1- S^{[\Box]}(\bm{x}_\perp,\bm{y}_\perp)$:
\begin{align}
{\cal F}^{[\Box]}(\bm{k}_\perp, \bm{\Delta}_\perp) & = \frac{4N_c}{\langle P | P\rangle} \int \frac{d^2 \bm{x}_\perp d^2 \bm{y}_\perp}{(2\pi)^2} \, 
e^{-ik\cdot (\bm{x}_\perp-\bm{y}_\perp)} e^{i\Delta \cdot (\bm{x}_\perp+\bm{y}_\perp)/2} 
\langle {P^\prime} \vert\; 1 - S^{[\Box]}(\bm{x}_\perp,\bm{y}_\perp)  \; \vert {P} \rangle \nonumber\\
& =  \frac{4N_c}{\langle P | P\rangle} \int \frac{d^2 \bm{r}_\perp d^2 \bm{b}_\perp}{(2\pi)^2} \, 
e^{-i\bm{k}_\perp\cdot \bm{r}_\perp} e^{i\bm{\Delta}_\perp \cdot \bm{b}_\perp} 
\langle {P^\prime} \vert\; 1- S^{[\Box]}(\bm{b}_\perp-\frac{\bm{r}_\perp}{2},\bm{b}_\perp+\frac{\bm{r}_\perp}{2})  \; \vert {P} \rangle. 
\end{align}

\begin{figure}[htb]
	\centering
	\includegraphics[width=6.5cm]{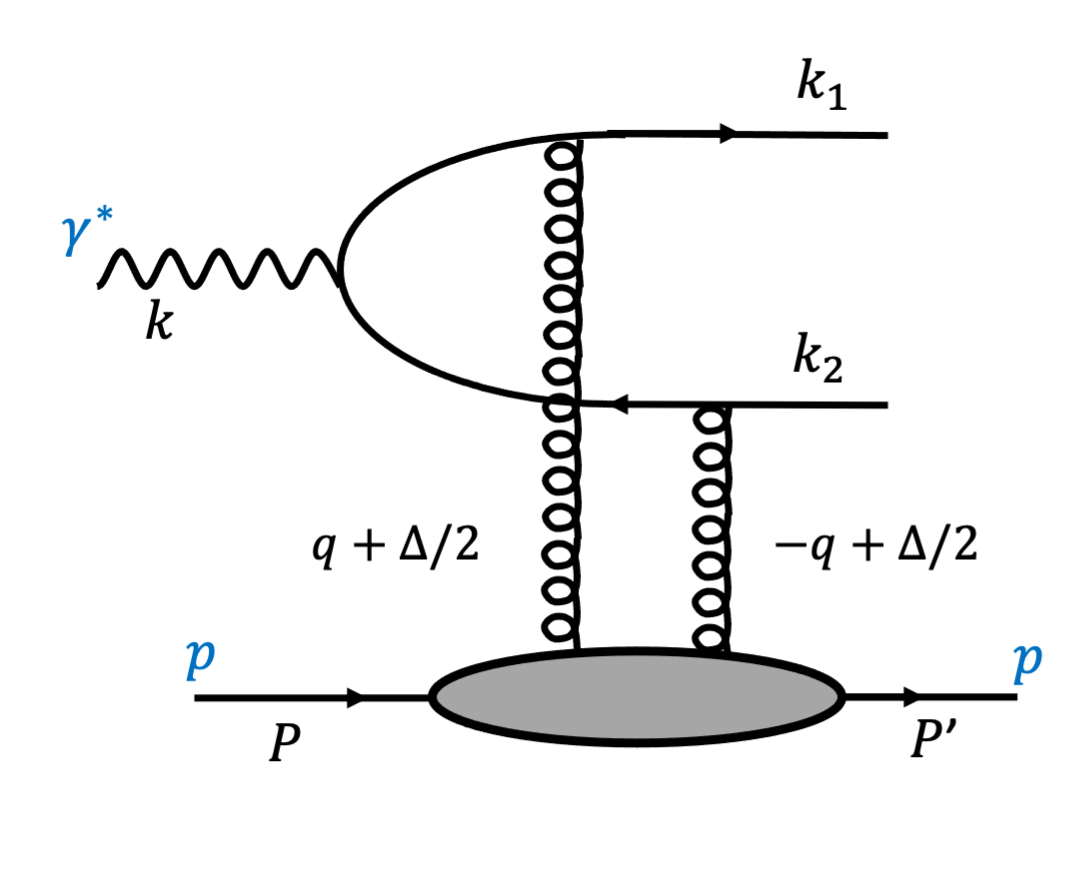}
	\caption{\label{Diagram} One of the leading order diagrams of diffractive dijet production in $ep$ collisions.}
\end{figure}

Following Ref.\ \cite{Hatta:2016dxp} the cross section of diffractive dijet production in electron-proton collisions is expressed in terms of ${\cal F}^{[\Box]}$ as
\begin{align}
\frac{d\sigma}{dy_1 dy_2 d^2 \bm{k}_{1\perp} d^2\bm{k}_{2\perp}} \propto \int d^2 \bm{q}_\perp d^2 \bm{q}^\prime_\perp  {\cal F}^{[\Box]}(\bm{q}_\perp,\bm{\Delta}_\perp) {\cal F}^{[\Box]}(\bm{q}^\prime_\perp,\bm{\Delta}_\perp) {\cal A}(\bm{K}_\perp, \bm{q}_\perp, \bm{q}^\prime_\perp,\epsilon_f^2),
\end{align}
for a particular amplitude function ${\cal A}$. Here $K_\perp$ is the transverse part of $K=(k_1-k_2)/2$, where $k_i$ denotes the momentum of jet $i$, $\Delta_\perp$ is the transverse part of $\Delta = k_1+k_2$, $y_i$ is the rapidity of jet $i$, and $\epsilon_f^2 = z(1-z)Q^2$, with $z$ is the momentum fraction of one of the two jets. One considers this process in the so-called correlation limit: $\Delta_\perp \ll K_\perp$, where 
$K_\perp$ sets the hard scale, allowing to also consider the photoproduction ($Q^2=0$) case. One of the leading order diagrams of diffractive dijet production in $ep$ collisions in this kinematic regime is shown in Fig. \ref{Diagram}. In this exclusive process the transverse momentum of the jet pair gives a handle on the $\bm{\Delta}_\perp$ momentum, even if the off-forwardness of the struck nucleon or nucleus itself is not measured. More details of the cross section calculation will be given below, but first we will discuss the model for the Wilson loop GTMD $G^{[\Box]}$ and the corresponding ${\cal F}^{[\Box]}$.

\section{Model for the unpolarized gluon GTMD at small $x$\label{model}}

In the process considered the incoming virtual photon splits into a quark-antiquark dipole pair that interacts with the proton or nucleus.\footnote{Note that for sufficiently high center of mass energy of the scattering this dipole picture can be reconciled with the target having a large $P^+$ momentum which, as we discussed, was required to consider the impact parameter dependence w.r.t.\ to a sufficiently well determined center $R_\perp=0$ of the target. The appropriate frame is referred to as the dipole frame \cite{Mueller:2001fv}. In addition, for large dipole sizes the impact parameter should also be defined w.r.t.\ the center of momentum of the dipole \cite{Bartels:2003yj}, but that is not relevant for the small dipole sizes considered here.}.
The large jet transverse momentum, or equivalently large $K_\perp$, the typical size $r_\perp$ of the dipole will be small. At small enough $x$ even small dipoles will have multiple interactions with the target. In the saturation regime at small $x$ one often employs the McLerran-Venugopalan (MV) model for the dipole scattering amplitude \cite{McLerran:1993ni,McLerran:1993ka,McLerran:1994vd,Gelis:2001da,Gelis:2010nm}:  
\begin{align}
\left\langle \frac{S^{[\Box]}(\bm{x}_\perp,\bm{y}_\perp) + S^{[\Box]\dagger}(\bm{x}_\perp,\bm{y}_\perp)}{2} \right\rangle_C
= \exp \left(-\frac{1}{4} {r}_\perp^2 Q_s^2 \ln \left[\frac{1}{{r_\perp^2}\Lambda^2}+e\right]\right),
\label{MVmodel}
\end{align}
where the subscript $C$ indicates that an average over the color configuration of the target is taken, $\Lambda$ denotes the QCD scale, and $e$ is the natural number. For an infinitely large nuclear target the saturation scale $Q_s$ is only a function of $x$. As a consequence, in that case the MV model expression applies to the forward scattering case and it is only a function of $\bm{r}_\perp=\bm{y}_\perp-\bm{x}_\perp$ due to translational invariance ($r_\perp^2 = |\bm{r}_\perp^2|$). For finite nuclei at small $x$, a dependence of $Q_s$ on impact parameter is often considered, see e.g.\ \cite{Mueller:1989st,GolecBiernat:2003ym,Iancu:2017fzn}. The $b_\perp (=|\bm{b}_\perp|)$ dependence of $Q_s$ is usually implemented as $Q_s^2(x,b_\perp) \equiv Q_s^2(x) T_A(b_\perp)$, where $T_A(b_\perp)$ is the nuclear profile function or nuclear thickness function that describes the distribution of nuclear matter inside a nucleus integrated over the $z$ component of $\bm{b}$ \cite{Iancu:2017fzn}. Here $T_A(b_\perp)$ scales with $A^{1/3}$. 
For scattering off a proton at very small $x$ that is described by the CGC, one can similarly introduce a profile function. 
To be specific, we will consider
\begin{align}
Q_s^2({b}_\perp) = \frac{4\pi \alpha_s^2 C_F}{N_c} T_p({b}_\perp),
\end{align}
with a Gaussian ${b}_\perp$ profile \cite{Salazar:2019ncp}: 
\begin{equation}
T_p(b_\perp) = \exp\left(-b_\perp^2/(2 R_p^2)\right),
\label{Proton}
\end{equation}
where $R_p$ is the gluonic radius of the proton for which we will take the value $R_p=0.5 \ \text{fm}$, such that $1/R_p \approx 0.4$ GeV.

Using Eq.\ (\ref{MVmodel}) with this $Q_s^2({b}_\perp)$ implies automatically nonzero off-forwardness, even if one is considering only diagonal expectation values.
Furthermore, by identifying (and implicitly absorbing the lightcone volume factor $2P^+ \int db^-$ of $\langle P | P\rangle$ in the process)
\begin{align}
\langle S^{[\Box]}(\bm{b}_\perp,\bm{r}_\perp) \rangle_C = \langle P^+, \bm{R}_\perp = 0 \vert  S^{[\Box]}(\bm{b}_\perp,\bm{r}_\perp) \vert P^+, \bm{R}_\perp = 0 \rangle,
\end{align}
we arrive at the following expression for the GTMD: 
\begin{align}
G^{[\Box]}(\bm{k}_\perp, \bm{\Delta}_\perp) & = 4N_c \int \frac{d^2 \bm{r}_\perp d^2 \bm{b}_\perp}{(2\pi)^2} \, 
e^{-i\bm{k}_\perp\cdot \bm{r}_\perp} e^{i\bm{\Delta}_\perp \cdot \bm{b}_\perp} \exp \left(-\frac{1}{4} {r}_\perp^2 Q_s^2(b_\perp) \ln \left[\frac{1}{{r}_\perp^2\Lambda^2} +e \right]\right).
\label{ourmodel}
\end{align}
This becomes the standard MV model expression for the gluon TMD in the limit $R_p \to \infty$ and $\Delta_\perp \to 0$. 
We expect this model expression to be applicable as long as the typical $b_\perp$ values probed are larger than the typical dipole sizes. Therefore, we will restrict application of the model to the region $\Delta_\perp \simorder 1$ GeV, which is consistent with the correlation limit, because well-defined jets will have transverse momenta of at least a few GeV. In practice, higher $\Delta_\perp$ will hardly matter, as will be seen (cf.\ Fig.\ \ref{AT_k}).

In \cite{Hagiwara:2016kam,Hagiwara:2017fye} Gaussian weighting factors $e^{-\epsilon_r r_\perp^2}$ and $e^{-\epsilon_b b_\perp^2}$ are introduced as cut offs. This will cut out the regions where the $q\bar{q}$ dipole does not overlap with the target or its size becomes large compared to the target size, where the model should not be applicable. Here we will only introduce $e^{-\epsilon_r r_\perp^2}$, as we found that there is actually no need for $e^{-\epsilon_b b_\perp^2}$ when considering ${\cal F}^{[\Box]}$. To be specific, in order to fit the model to H1 data, we will introduce two free parameters $\{\epsilon_r, \chi\}$ in the model in the following way: 
\begin{align}
{\cal F}^{[\Box]}(\bm{k}_\perp, \bm{\Delta}_\perp) = 4N_c \int \frac{d^2 \bm{r}_\perp d^2 \bm{b}_\perp}{(2\pi)^2} \, 
e^{-i\bm{k}_\perp\cdot \bm{r}_\perp} e^{i\bm{\Delta}_\perp \cdot \bm{b}_\perp} \, e^{-\epsilon_r r_\perp^2} \, \left[1- 
\exp \left(-\frac{1}{4} {r}_\perp^2 \chi Q_s^2(b_\perp) \ln \left[\frac{1}{{r}_\perp^2\Lambda^2} +e\right]\right)\right].
\label{modelFbox}
\end{align}
We will consider a fixed value $\Lambda = 0.24\, \mathrm{GeV}$ and consider $N_f=4$ for the number of active flavors. The fitted $\chi$ value can be viewed as determining the $x$ value of the model through the $x$ dependence of the saturation scale. In applications of the MV model the saturation scale is usually taken to be $Q_s^2(x)= A^{1/3} (3\cdot 10^{-4}/x)^{0.3}~[{\rm GeV}^2]$, that stems from the GBW (geometric scaling) description of the inclusive DIS data from HERA \cite{GolecBiernat:1998js}. Equating this expression with $\chi Q_s^2(b_\perp=0_\perp) = 0.5 \chi A^{1/3}~[{\rm GeV}^2]$, one finds $\chi = 2 (3\cdot 10^{-4}/x)^{0.3}$. 

Next one can expand equation (\ref{modelFbox}) as
\begin{equation}
{\cal F}^{[\Box]}(\bm{k}_\perp, \bm{\Delta}_\perp)
	= {\cal F}_0^{[\Box]}(k_\perp, \Delta_\perp) + 2 {\cal F}_2^{[\Box]}(k_\perp, \Delta_\perp) \cos 2\theta_{k\Delta} + ...,
	\label{Expand_Fbox}
\end{equation}
where $\theta_{k\Delta}$ denotes the angle between $\bm{k}_\perp$ and $\bm{\Delta}_\perp$. 
The contribution of the elliptic part ${\cal F}_2^{[\Box]}$, and even more so of the higher order harmonics, to the cross section will be small compared to the angular independent part ${\cal F}_0^{[\Box]}$. Therefore, we will only retain the latter. 

\begin{figure}[htb]
	\centering
		\includegraphics[width=6.5cm]{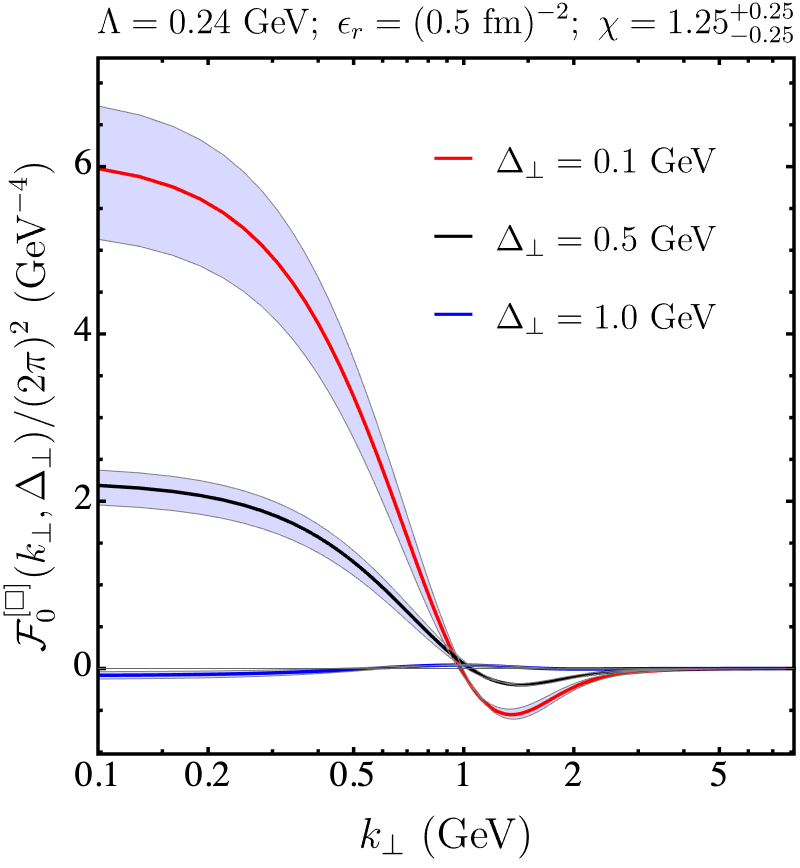}
		\caption{\label{F0_k} The function ${\cal F}_0^{[\Box]}/(2\pi)^2$ as a function of the transverse momentum $k_\perp$ for three different values of $\Delta_\perp$ for the choice $\epsilon_r=(0.5\, {\rm fm})^{-2}$. The curves are for $\chi=1.25$ and the bands around them correspond to $\chi$ in the range from 1.0 to 1.5, where larger $\chi$ yields larger results.}
\end{figure}

In Fig.\ \ref{F0_k} we show the function ${\cal F}_0^{[\Box]}$ for various parameters choices and ranges that we will consider below.

\section{Exclusive Diffractive dijet production cross sections \label{crosssection}}

The cross section for the diffractive dijet production process $\gamma^* p(A) \rightarrow q \bar{q} p(A)$ can be calculated 
at leading order (LO) by combining two steps: 1) the incoming virtual photon which splits into a quark-antiquark dipole pair and 2) the interaction of the pair with the proton or nucleus via two-gluon exchange. The LO of the virtual photon light cone wave function with virtuality $Q$ is discussed in many papers, such as \cite{Beuf:2011xd,Altinoluk:2015dpi,Hanninen:2017ddy} which we follow to arrive at the cross section expressions: 
\begin{eqnarray}
\frac{d\sigma^{\gamma^* p}_{T,L}}{dz_1 dz_2 d^2 \bm{K}_\perp d^2 \boldsymbol{\Delta}_\perp}
 =  \frac{1}{2 (2 \pi)^5 z_1 z_2} \delta(z_1 + z_2 - 1)\sum_{\beta_i h_i}\left| 
 \left\langle
 \mathcal{M}_{q \bar{q}}
 \right\rangle_C
 \right|^2,
\end{eqnarray}
where $ \mathcal{M}_{q \bar{q}}$ denotes the amplitude of this process, $z_{1,2}=k_{1,2}^+/k^+$ are the outgoing quark and antiquark longitudinal momentum fraction w.r.t.\ the virtual photon longitudinal momentum, and the sum is over color indices $\beta_i$ and quark helicities $h_i$. For the case of a transverse photon the amplitude is given by:
\begin{eqnarray}
\mathcal{M}^T_{q \bar{q}} 
&=& 
e e_f \sqrt{z_1 z_2} 
\left[ z_2 - z_1 - 2 h_0 \lambda \right]  \delta_{h_0, -h_1}\
	\int d^2 \bm{q}_\perp 
		\int \frac{d^2 \bm{r}_\perp\, d^2 \bm{b}_\perp}{(2\pi)^2}
	e^{-i\bm{b}_\perp \cdot \boldsymbol{\Delta}_\perp}
	e^{-i\bm{r}_\perp \cdot \bm{q}_\perp} 
	\nonumber \\
	&\times&  \left[ U_ {\beta_0 \beta_1}^{[\Box]}(\bm{b}_\perp-\frac{\bm{r}_\perp}{2},\bm{b}_\perp+\frac{\bm{r}_\perp}{2}) -  \delta_{\beta_0 \beta_1} \right] \frac{\boldsymbol{\epsilon}_\lambda \cdot  \left(\bm{K}_\perp -\bm{q}_\perp \right) }{z_1 z_2 Q^2 + \left(\bm{K}_\perp -\bm{q}_\perp \right)^2},
\end{eqnarray}
and for a longitudinal photon by:
\begin{eqnarray}
	\mathcal{M}^L_{q \bar{q}} 
	&=& 
- 2 e e_f \sqrt{z_1 z_2}  \delta_{h_0, -h_1} 
	\int d^2 \bm{q}_\perp 
\int \frac{d^2 \bm{r}_\perp\, d^2 \bm{b}_\perp}{(2\pi)^2}
e^{-i\bm{b}_\perp \cdot \boldsymbol{\Delta}_\perp}
e^{-i\bm{r}_\perp \cdot \bm{q}_\perp} \nonumber \\
	&\times& \left[ U_ {\beta_0 \beta_1}^{[\Box]}(\bm{b}_\perp-\frac{\bm{r}_\perp}{2},\bm{b}_\perp+\frac{\bm{r}_\perp}{2}) -  \delta_{\beta_0 \beta_1} \right] 
  \frac{z_1 z_2 Q}{z_1 z_2 Q^2 + (\bm{K}_\perp - \bm{q}_\perp)^2},
\end{eqnarray}
where $\boldsymbol{\epsilon}_\lambda$ denotes the polarization vector of a photon with helicity $\lambda$. 
After averaging over color and photon polarization, and after summing over quark helicities \cite{Altinoluk:2015dpi}, we arrive at:
\begin{eqnarray}
\frac{d\sigma^{\gamma^* p}_T}{dz_1 dz_2 d^2  \bm{K}_\perp d^2 \boldsymbol{\Delta}_\perp} 
= && \frac{\alpha_{em}}{8(2\pi)^4 N_c} \sum_f e_f^2 \  \delta(z_1+z_2-1 )   \left[ z_1^2 +z_2^2 \right] 	\int d^2 \bm{q}_\perp \ \int d^2 \bm{q}_\perp^\prime \
{\cal F}^{[\Box]}(\bm{q}_\perp, \boldsymbol{\Delta}_\perp)  \nonumber \\
 && \times  {\cal F}^{[\Box]}(\bm{q}_\perp^\prime, \boldsymbol{\Delta}_\perp) 
\left[\frac{ \left(\bm{K}_\perp -\bm{q}_\perp \right) }{z_1 z_2 Q^2 + \left(\bm{K}_\perp -\bm{q}_\perp \right)^2}\right]
 \cdot  
\left[\frac{ \left(\bm{K}_\perp -\bm{q}_\perp^\prime \right) }{z_1 z_2 Q^2 + \left(\bm{K}_\perp -\bm{q}_\perp^\prime \right)^2}\right],
	 \label{Cross_T}
\end{eqnarray}
and 
\begin{eqnarray}
	\frac{d\sigma^{\gamma^* p}_L}{dz_1 dz_2 d^2 \bm{K}_\perp d^2 \boldsymbol{\Delta}_\perp} 
	= &&
\frac{\alpha_{em}}{2(2\pi)^4 N_c} \sum_f e_f^2 \ \delta(z_1+z_2-1 )  \ z_1^2 z_2^2 Q^2	\int d^2 \bm{q}_\perp \ 	\int d^2 \bm{q}_\perp^\prime \
{\cal F}^{[\Box]}(\bm{q}_\perp, \boldsymbol{\Delta}_\perp)  \nonumber \\
	&& \times {\cal F}^{[\Box]}(\bm{q}_\perp^\prime, \boldsymbol{\Delta}_\perp) \frac{1 }{z_1 z_2 Q^2 + \left(\bm{K}_\perp -\bm{q}_\perp \right)^2}
	 \frac{1}{z_1 z_2 Q^2 + \left(\bm{K}_\perp -\bm{q}_\perp^\prime \right)^2}.
	 \label{Cross_L}
\end{eqnarray}
As discussed earlier we will only consider the angular independent part ${\cal F}_0^{[\Box]}$. 

We note that since there is an average over the color configurations of the target, the cross section will scale as $N_c$, the sum over colors of the quark-antiquark pair. Since we express the cross section in terms of GTMDs that themselves scale as $N_c$, the above expressions have an overall $1/N_c$ factor, rather than $N_c$ like in e.g.\ \cite{Altinoluk:2015dpi}. But the results are consistent with each other.

In order to relate $\sigma^{\gamma^* p}$ to the diffractive dijet cross section $\sigma^{ep}$ for electron-proton collisions in DIS in HERA experiments, one can use \cite{Altinoluk:2015dpi}
\begin{eqnarray}
	\frac{d \sigma^{ep}}{dx \ dQ^2 } = \frac{\alpha_{em}}{\pi x Q^2 }
	\left[
	\left(
	1-y+\frac{y^2}{2}
	\right)  \sigma^{\gamma^* p}_T
	+ \left(
	1-y
	\right) \sigma^{\gamma^* p}_L
	\right],
	\label{dSep}
\end{eqnarray}
with $y=Q^2/sx$ and $\sqrt{s}$ is the center of mass energy of $ep$ collision, which for the H1 data to be discussed was 319 GeV.
We will fit the resulting expression to the high $Q^2$ electroproduction data of H1 and make predictions for high $Q^2$ electroproduction at EIC. We will also make predictions for photoproduction, which we now discuss.
 
\subsection{Photoproduction: $Q^2 =0 $}

\begin{figure}[htb]
	\centering
	\includegraphics[width=6.5cm]{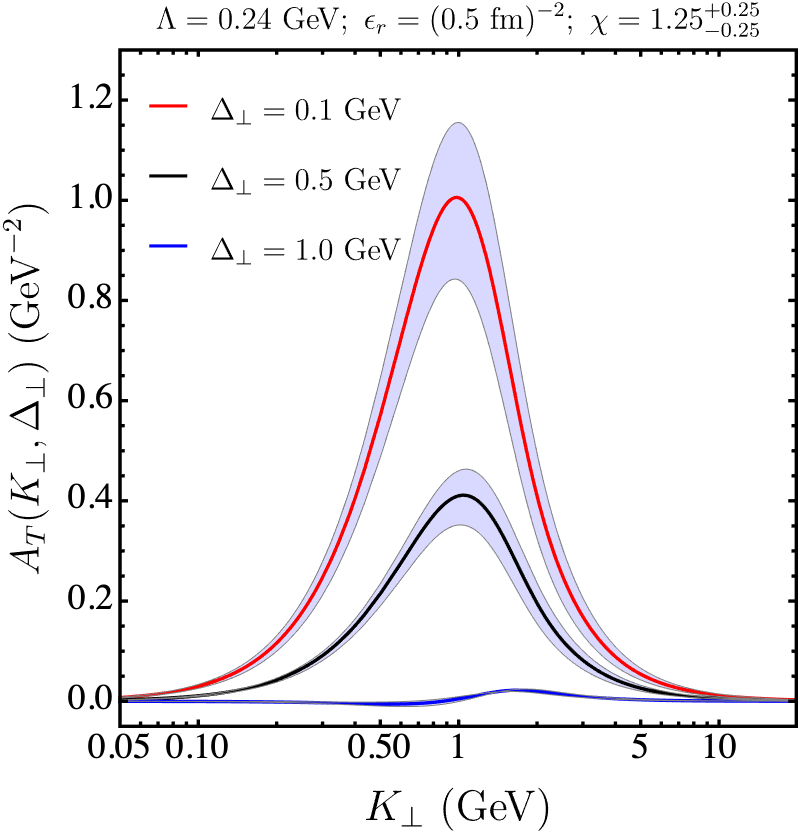}
	\caption{\label{AT_k} The function $A_T (K_\perp, \Delta_\perp)$ as a function of $K_\perp$ for three different values of $\Delta_\perp$.}
\end{figure}

For the case of $Q^2=0$ the expression for $d\sigma^{\gamma^* p}_T$ can be used to arrive at:
\begin{eqnarray}
\frac{d\sigma^{\gamma p}}{dz_1 dz_2 d^2  \bm{K}_\perp d^2 \boldsymbol{\Delta}_\perp} 
= && \frac{\alpha_{em}}{8(2\pi)^4 N_c} \sum_f e_f^2 \  \delta(z_1+z_2-1 )   \left[ z_1^2 +z_2^2 \right] 	\int d^2 \bm{q}_\perp \ \int d^2 \bm{q}_\perp^\prime \
{\cal F}^{[\Box]}(\bm{q}_\perp, \boldsymbol{\Delta}_\perp)  \nonumber \\
 && \times  {\cal F}^{[\Box]}(\bm{q}_\perp^\prime, \boldsymbol{\Delta}_\perp) 
\left[\frac{ \left(\bm{K}_\perp -\bm{q}_\perp \right) }{\left(\bm{K}_\perp -\bm{q}_\perp \right)^2}\right]
 \cdot  
\left[\frac{ \left(\bm{K}_\perp -\bm{q}_\perp^\prime \right) }{\left(\bm{K}_\perp -\bm{q}_\perp^\prime \right)^2}\right].
	 \label{Cross_gamma}
\end{eqnarray}
The integrations over the angles of $\bm{q}_\perp$ and $\bm{q}_\perp^\prime$ can be calculated analytically to arrive at \cite{Hagiwara:2017fye,Pelicer:2018clw}
\begin{eqnarray}
	\int d^2 \bm{q}_\perp 
\frac{  {\cal F}_0^{[\Box]} (q_\perp, {\Delta}_\perp)  \left(\bm{K}_\perp -\bm{q}_\perp \right) }{ \left(\bm{K}_\perp -\bm{q}_\perp \right)^2} =  \frac{(2 \pi)^3 \bm{K}_\perp }{K_\perp^2} A_T (K_\perp, \Delta_\perp),
\label{int_q}
\end{eqnarray}
with 
\begin{eqnarray}
A_T (K_\perp, \Delta_\perp) = \frac{1}{(2\pi)^2} \int_0^{K_\perp} dq_\perp \, q_\perp \, {\cal F}_0^{[\Box]}(q_\perp, {\Delta}_\perp).
\end{eqnarray}
In Fig.\ \ref{AT_k} we show the function $A_T (K_\perp, \Delta_\perp)$ as a function of $K_\perp$ for the same parameters choices and ranges as in Fig.\ \ref{F0_k}. It shows that the function is already very small when $\Delta_\perp = 1$ GeV. 

The $\gamma p$ differential cross section can thus be written as:
\begin{eqnarray}
		\frac{d\sigma^{\gamma p}}{dz_1 dz_2 dK_\perp d \Delta_\perp^2} 
		= && \frac{(2 \pi)^4\alpha_{em}}{16 N_c} \sum_f e_f^2 \  \delta(z_1+z_2-1 )   \left[ z_1^2 +z_2^2 \right] \frac{A^2_T (K_\perp, \Delta_\perp)}{K_\perp}.
		\label{Cross_gamma2}
	\end{eqnarray}

\subsection{Electroproduction: $Q^2 >  0$}

For the case of $Q^2 >  0 $ the cross section receives contributions from both the transverse and the longitudinal polarization states of the photon. The integrations over the angles of $\bm{q}_\perp$ and $\bm{q}_\perp^\prime$ can again be calculated analytically to arrive for the transverse part in Eq.\ (\ref{Cross_T}) at
\begin{equation}
	\int d^2 \bm{q}_\perp \mathcal{F}^{[\Box]}_0 (q_\perp,{\Delta_\perp}) 
	\left[\frac{ \left(\bm{K}_\perp -\bm{q}_\perp \right) }{z_1 z_2 Q^2 + \left(\bm{K}_\perp -\bm{q}_\perp \right)^2}\right] 
	=
	\frac{(2 \pi)^3 \bm{K}_\perp}{K_\perp^2} {\cal A}_{\text{T}} (K_\perp, \Delta_\perp, z_i, Q)
\end{equation}
with
\begin{equation} 
	{\cal A}_{\text{T}} (K_\perp, \Delta_\perp, z_i, Q) = \frac{1}{(2 \pi)^2}
	\int_0^\infty dq_\perp 
	\ \frac{q_\perp \ \mathcal{F}^{[\Box]}_0 (q_\perp,{\Delta_\perp}) }{2}  \left[ 1+ \frac{K_\perp^2 - q_\perp^2 - z_1 z_2 Q^2 }{\sqrt{(K_\perp^2+q_\perp^2+z_1 z_2 Q^2)^2-(2 K_\perp q_\perp)^2}} \right],
	\label{A_Transv}
\end{equation}
while for the longitudinal part in Eq.\ (\ref{Cross_L}) can be evaluated to be
\begin{equation}
	\int d^2 \bm{q}_\perp\mathcal{F}^{[\Box]}_0 (q_\perp,{\Delta_\perp}) 
	\left[\frac{ Q }{z_1 z_2 Q^2 + \left(\bm{K}_\perp -\bm{q}_\perp \right)^2}\right] 
	= \frac{(2 \pi)^3 K_\perp }{K_\perp^2}  {\cal A}_{\text{L}} (K_\perp, \Delta_\perp, z_i, Q)
\end{equation}
with
\begin{equation}
	{\cal A}_{\text{L}} (K_\perp, \Delta_\perp, z_i, Q) = \frac{1}{(2 \pi)^2} \int_0^{\infty}  \ dq_\perp \ q_\perp \  \mathcal{F}^{[\Box]}_0 (q_\perp,{\Delta_\perp})
	\left[\frac{ K_\perp Q }
	{ \sqrt{( K_\perp^2 +q_\perp^2 + z_1 z_2 Q^2 )^2 - (2 K_\perp q_\perp)^2}
	} \right].
\label{A_Long}
\end{equation}
Therefore, the $\gamma^* p$ differential cross section for $Q^2 > 0$ can be expressed in terms of ${\cal A}_{\text{T}}$ and ${\cal A}_{\text{L}}$ as
\begin{equation}
	\frac{d\sigma^{\gamma^* p}_L}{dz_1 dz_2 d K_\perp  d\Delta_\perp^2} 
	=  \frac{(2 \pi)^4 \alpha_{em}}{16 N_c} \sum_f e_f^2 \  \delta(z_1+z_2-1 )   \left[ z_1^2 +z_2^2 \right] \frac{{\cal A}^2_{\text{T}} (K_\perp, \Delta_\perp, z_i, Q) }{K_\perp}
	\label{ATransv}
\end{equation}
and 
\begin{equation}
	\frac{d\sigma^{\gamma^* p}_L}{dz_1 dz_2 d K_\perp  d\Delta_\perp^2} 
	=   \frac{(2 \pi)^4 \alpha_{em}}{4 N_c} \sum_f e_f^2 \  \delta(z_1+z_2-1 )  z_1^2 z_2^2  \frac{{\cal A}^2_{\text{L}} (K_\perp, \Delta_\perp, z_i, Q) }{K_\perp}.
	\label{ALong}
\end{equation}

\begin{figure}[htb]
	\begin{minipage}{0.45\textwidth}
		\centering
		\includegraphics[width=6.5cm]{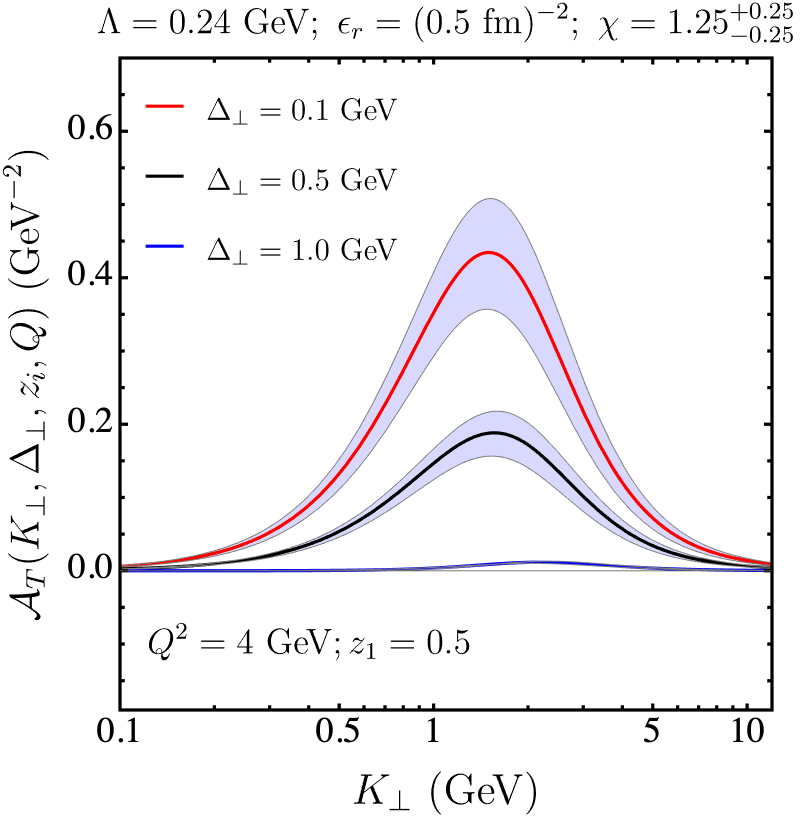}
		\caption{\label{AT_Full} The function ${\cal A}_{\text{T}}$ as a function of $K_\perp$ from Eq. (\ref{A_Transv}) for three different values of $\Delta_\perp$, $Q^2 = 4 \  \text{GeV}^2$, and $z_1=z_2=0.5$.}
	\end{minipage} \hfill
	\begin{minipage}{0.45\textwidth}
		\centering
		\includegraphics[width=6.5cm]{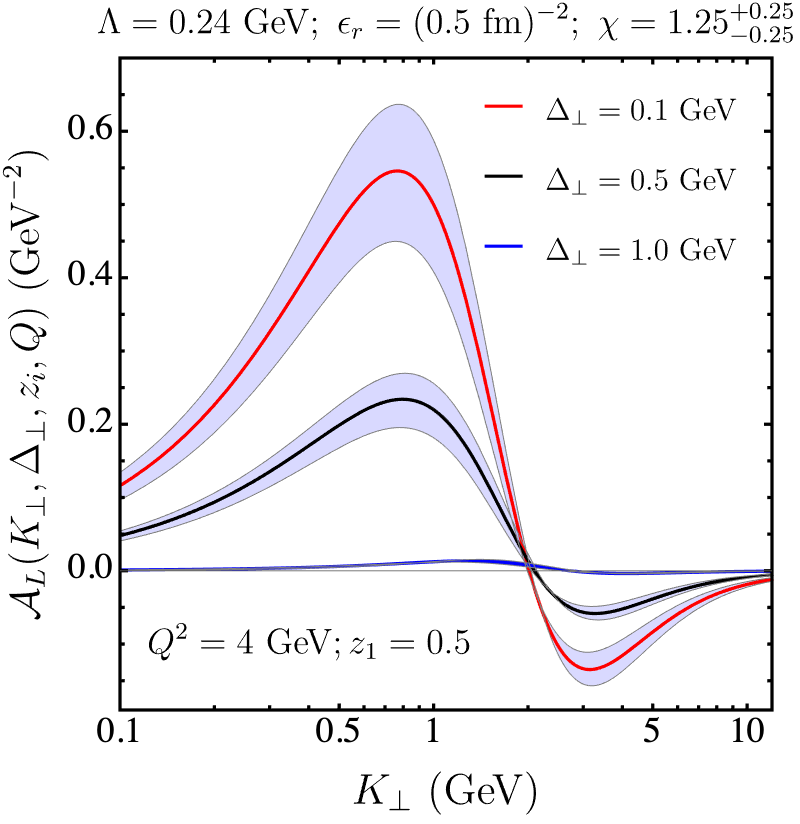}
		\caption{\label{AL_Full} The function ${\cal A}_{\text{L}}$ as a function of $K_\perp$ from Eq. (\ref{A_Long}) for three different values of $\Delta_\perp$, $Q^2 = 4 \  \text{GeV}^2$, and $z_1=z_2=0.5$.}
	\end{minipage} \hfill
\end{figure}

\begin{figure}[htb]
	\begin{minipage}{0.45\textwidth}
		\centering
		\includegraphics[width=6.5cm]{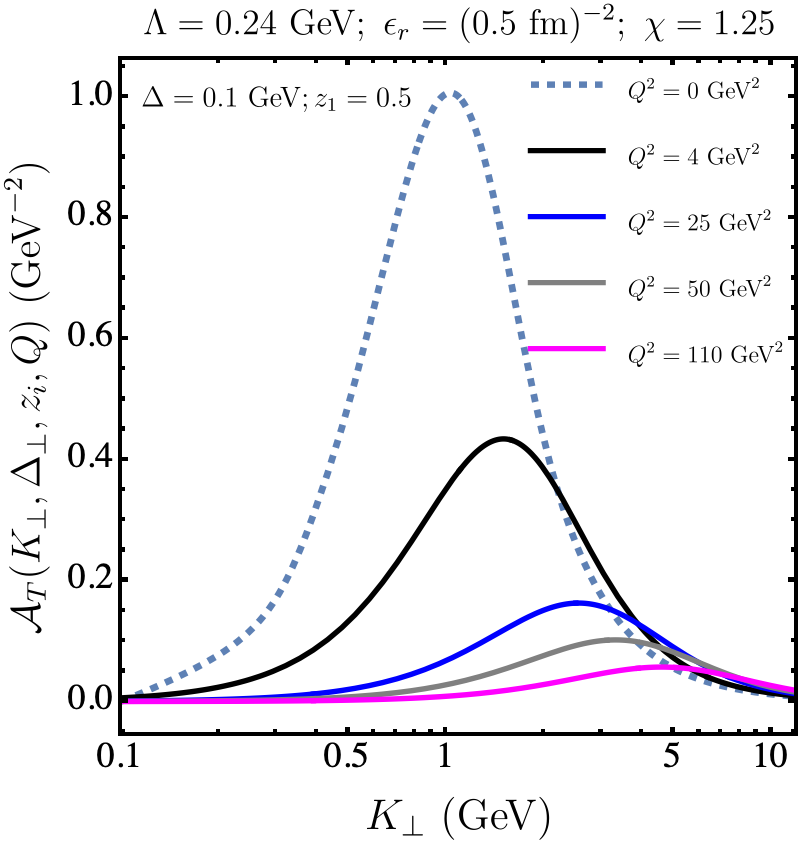}
		\caption{\label{AT_Q2} The function ${\cal A}_{\text{T}}$ as a function of $K_\perp$ from Eq. (\ref{A_Transv}) for five different values of $Q^2$, $\Delta_\perp=0.1 \ \text{GeV}$, and $z_1=z_2=0.5$.}
	\end{minipage} \hfill
	\begin{minipage}{0.45\textwidth}
		\centering
		\includegraphics[width=6.5cm]{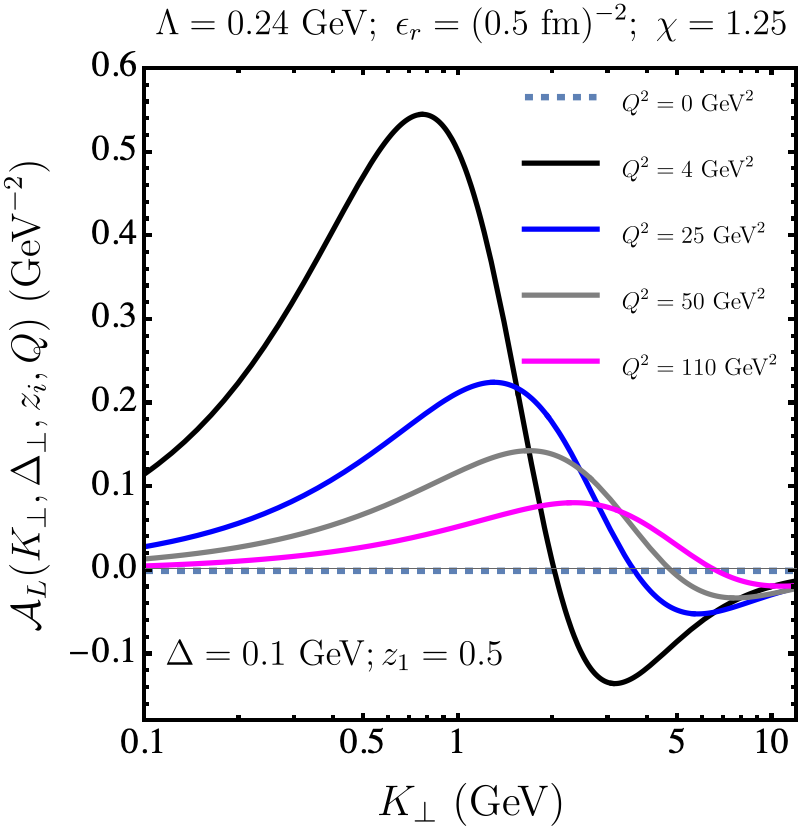}
		\caption{\label{AL_Q2}  The function ${\cal A}_{\text{L}}$ as a function of $K_\perp$ from Eq. (\ref{A_Long}) for five different values of $Q^2$, $\Delta_\perp=0.1 \ \text{GeV}$, and $z_1=z_2=0.5$.}
	\end{minipage} \hfill
\end{figure}

In Figs.\ \ref{AT_Full} and \ref{AL_Full} we show the functions ${\cal A}_{\text{T}}$ and ${\cal A}_{\text{L}}$ for the same parameter choices as in Fig.  \ref{F0_k} and  Fig. \ref{AT_k}, and for $Q^2= 4 \  \text{GeV}^2$ and $z_1=z_2=0.5$. It can be seen that the magnitude of ${\cal A}_{\text{L}}$ is larger than ${\cal A}_{\text{T}}$. However, the $z_1^2 z_2^2 $ term in front of ${\cal A}_{\text{L}}$ makes its contribution to the differential cross section much smaller as can be seen in the next section. In Figs.\ \ref{AT_Q2} and \ref{AL_Q2} we show ${\cal A}_{\text{T}}$ and ${\cal A}_{\text{L}}$ for different values of $Q^2$. 

\section{Model fit of H1 data\label{fits}}

\begin{figure}[htb]
	\begin{minipage}{0.45\textwidth}
		\centering
		\includegraphics[width=6.5cm]{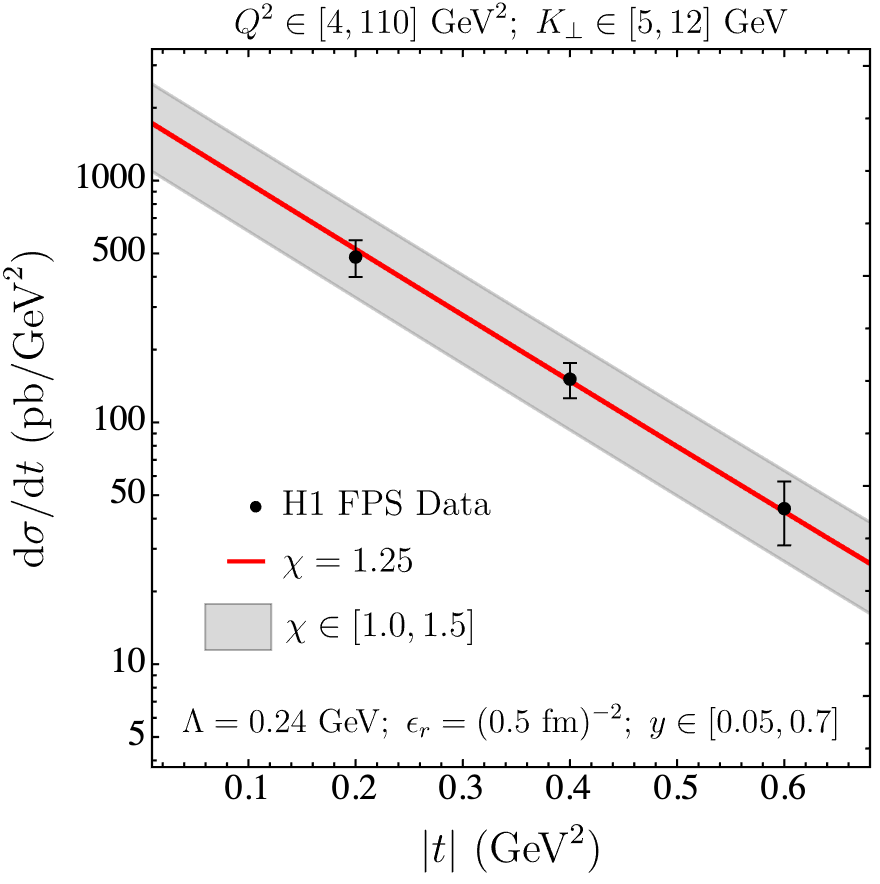}
		\caption{\label{dSdt_H1} Differential cross section as a function of $|t|$ for the model for the indicated parameter choices and ranges, and for the H1 FPS data including the total uncertainties $\delta_\text{tot}$.}
	\end{minipage} \hfill
	\begin{minipage}{0.45\textwidth}
		\centering
		\includegraphics[width=6.5cm]{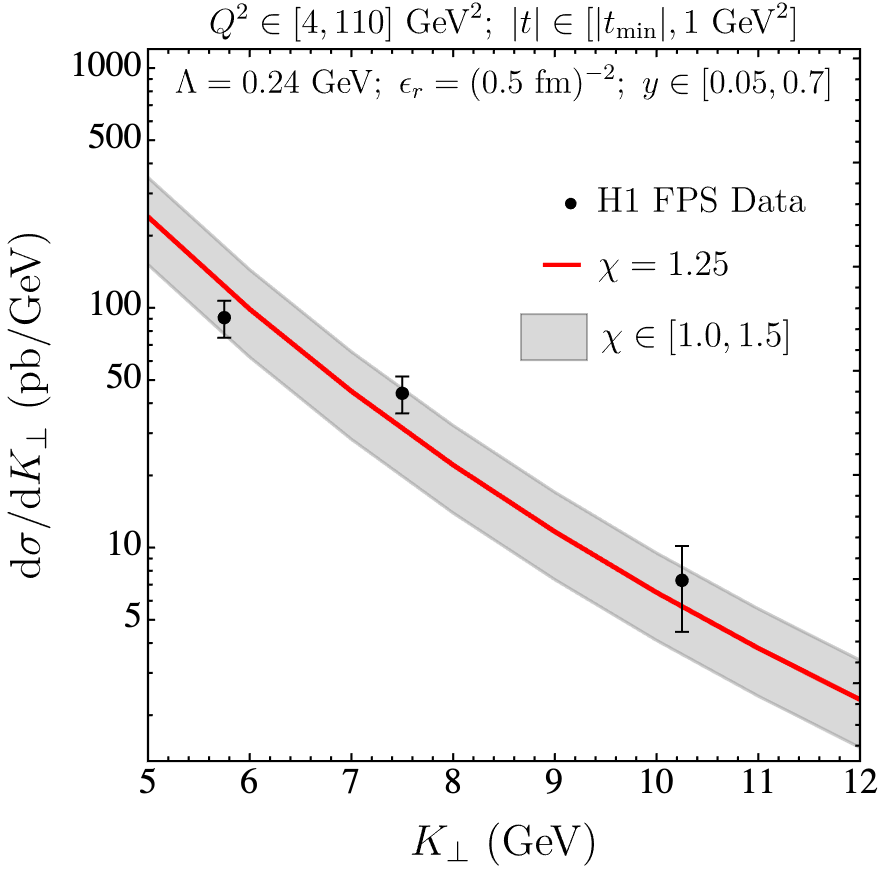}
		\caption{\label{dSdK_H1} Differential cross section as a function of $K_\perp \approx k_{1\perp}$ for the model for the indicated parameter choices and ranges ($t_{\min} \approx 0$), and for the H1 FPS data including the total uncertainties $\delta_\text{tot}$.}
	\end{minipage} \hfill
\end{figure}

The H1 and ZEUS experiments at HERA have studied the diffractive dijet process in a series of papers \cite{Aktas:2007bv,Aaron:2011mp,Abramowicz:2015vnu}. Here we focus on \cite{Aaron:2011mp} where data in the $Q^2$ range of $4-110\ {\rm GeV}^2$ was presented, for $y \in [0.05,0.7]$ and $|t| \leq 1\ {\rm GeV}^2$, that allows us to study the correlation limit region
\footnote{Strictly speaking, these H1 data are not purely exclusive diffractive dijet events to which our LO analysis applies. The events are required to have at least two jets, i.e.\ $ep \to ejjX'p$, and leave some room for additional particles for which a $q \bar{q} g$ final state would need to be included, such as in \cite{Boussarie:2019ero}. We will proceed under the assumption that the dijet contribution dominates and that corrections are $\alpha_s$ suppressed.}. Given this $Q^2$ range we consider the case of four flavors. The H1 cross sections for two central jets shown in Fig.\ \ref{dSdt_H1} are obtained by extrapolation to the range $| t_{\text{min}}| \leq |t| \leq 1 \ \text{GeV}^2$ in order to compare to earlier results \cite{Aktas:2007bv}. We fit our model to this extrapolation range. Here, $| t_{\text{min}}| $ is the minimum kinematically accessible value of $|t|$. In our case $|t|=\Delta_\perp^2$.

We first consider the data for the $t$-dependence of the differential cross section based on Eqs.\ (\ref{ATransv}) and (\ref{ALong}). We select $\epsilon_r=(0.5\, {\rm fm})^{-2}$ and find that $\chi=1.25$ can describe the data quite well, as shown in Fig.\ \ref{dSdt_H1}, which has a very clear $e^{-bt}$ dependence, with $b\approx 6\ {\rm GeV}^2$. The slope of the cross section as a function of $|t|$ is controlled by the proton profile in Eq.\ (\ref{Proton}). The H1 data description does not depend much on $\epsilon_r$, therefore, we have chosen $\epsilon_r=0.5 \ \text{fm}$ in line with the gluonic radius of the proton $R_p$.

We also display a band corresponding  to $\chi$ in the range from 1.0 to 1.5. This range is selected on the basis of the $K_\perp$-dependence of the differential cross section. The H1 data is actually presented as a function of the transverse momentum of one of the jets, which is large (in the range $5-12 \ {\rm GeV}$) and almost back-to-back with the other jet in the transverse plane, such that one can expect that $q_\perp \ll K_\perp$, although the values of $q_\perp$ are not included in \cite{Aaron:2011mp}. On the basis of this expectation we approximate $K_\perp=(k_{1\perp}-k_{2\perp})/2= k_{1\perp}-q_\perp/2 \approx k_{1\perp}$. The result is shown in Fig.\ \ref{dSdK_H1}. As can be seen the transverse momentum dependence does not show as clear a power law fall-off as the model curves, hence, the considerable uncertainty in the $\chi$ value. The contribution of the longitudinal part to the cross section is not very large, it is 12 \% at $K_\perp=12$ GeV and becomes smaller for smaller $K_\perp$, e.g.\ it is 3.5 \% for $K_\perp=5$ GeV. 

\begin{figure}[htb]
	\begin{minipage}{0.45\textwidth}
		\centering
		\includegraphics[width=6.5cm]{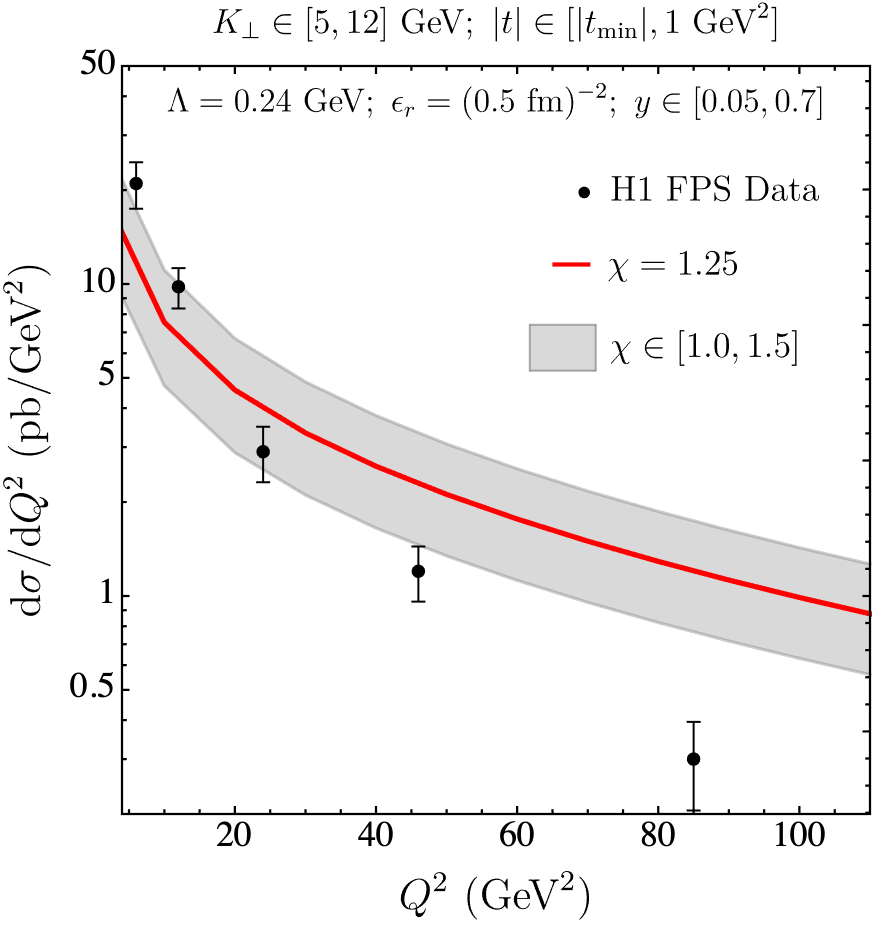}
		\caption{\label{dSdQ_H1} Differential cross section as a function of $Q^2$ for the model for the indicated parameter choices and ranges ($t_{\min} \approx 0$), and for the H1 FPS data including the total uncertainties $\delta_\text{tot}$.}
	\end{minipage} \hfill
\end{figure}

For the description of the H1 data we selected values for $\chi$ in the range $1.0-1.5$ with a central value $\chi=1.25$. One could relate these values using the GBW model expression for the saturation scale to corresponding $x$ values: $\chi = 2 (3\cdot 10^{-4}/x)^{0.3}$, such that $\chi=1.25\pm0.25$ corresponds to $x\sim (1-3) \cdot10^{-3}$. We consider such $\chi$ values to be acceptable and consistent with the typical $x$ values at which the MV model is generally considered to be applicable. Selecting a fixed $\chi$ value corresponds to selecting an average $x$ value. In Fig.\ \ref{dSdQ_H1} we show the $Q^2$ dependence of the model in comparison to the H1 data. It shows that for smaller $Q^2$, which corresponds to smaller $x$ values for given $y$ and $s$, a larger $\chi$ value is needed, whereas at larger $Q^2$, a smaller $\chi$ value is needed. Clearly a better description of the $Q^2$ dependence of the data could be obtained with an $x$-dependent $\chi$. However, since the integral of a curve through the central values of the H1 data as function of $Q^2$ is found to fall within the range of the $Q^2$ integral of the model, we proceed with the model with a fixed $\chi$, i.e.\ with an average $x$. We note that the H1 data spans an $x$ range from $5\cdot 10^{-5}$ to 0.02, giving a geometric mean of $10^{-3}$, which corresponds well with the $x$ values we obtain from the GBW model expression for the $\chi$ values considered. 

All in all, we conclude that the GTMD model allows for a fair description of the diffractive dijet production H1 data, for reasonable model parameters, in reasonable agreement with assumptions on the gluonic radius and the $x$ values the model applies to.

\section{Model predictions for EIC \label{predictions}}

In this final section we present predictions for exclusive diffractive dijet production at the EIC using our model. For leptoproduction we consider the range $3\leq K_\perp \leq 9\ {\rm GeV}$, because the center of mass energy of EIC will be lower than at HERA. At even lower $K_\perp$ we expect that jets cannot be resolved anymore and by selecting this range, we can consider the fixed flavor case with $N_f = 4$. We consider $Q^2 \in [1\ {\rm GeV}^2,K_\perp^2]$ rather than the fixed range $Q^2 \in [4,110]\  {\rm GeV}^2$ of HERA and also show the cases for $Q^2=K_\perp^2$ and $Q^2=4 K_\perp^2$ (here we expect smaller $Q^2$ to be better described by larger $\chi$ and vice versa).  
We also present model predictions for photoproduction. As expected, the cross section is much larger in this case.

\begin{figure}[htb]
	\begin{minipage}{0.45\textwidth}
		\centering
		\centering
		\includegraphics[width=6.5cm]{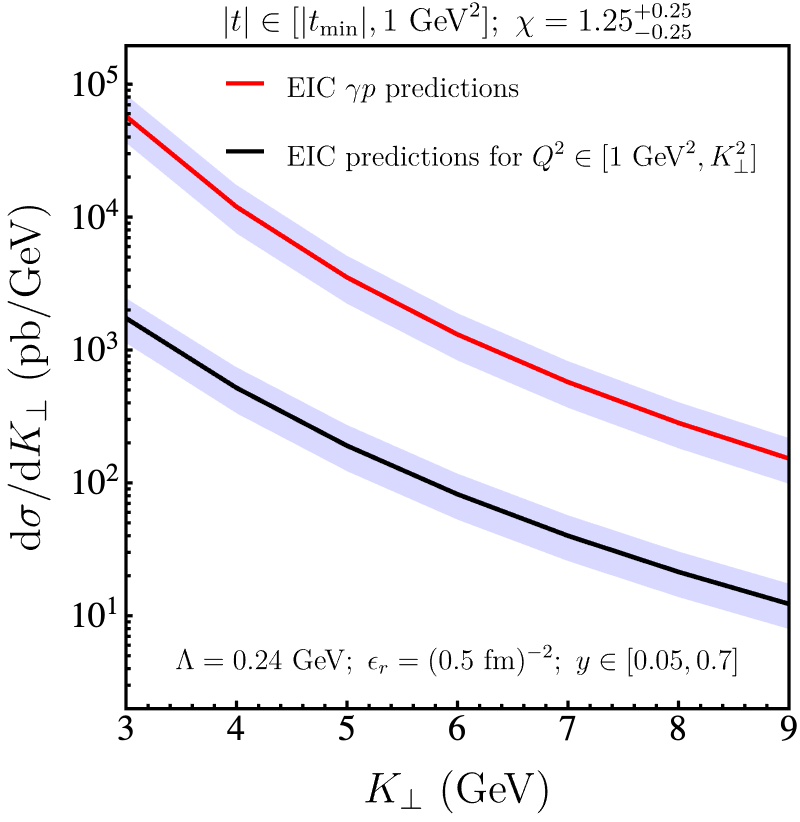}
		\caption{\label{EIC} Predictions for exclusive diffractive dijet production in $ep$ (black curve) and $\gamma p$ (red curve) collisions at the EIC.}
	\end{minipage} \hfill
	\begin{minipage}{0.45\textwidth}
		\centering
		\includegraphics[width=6.5cm]{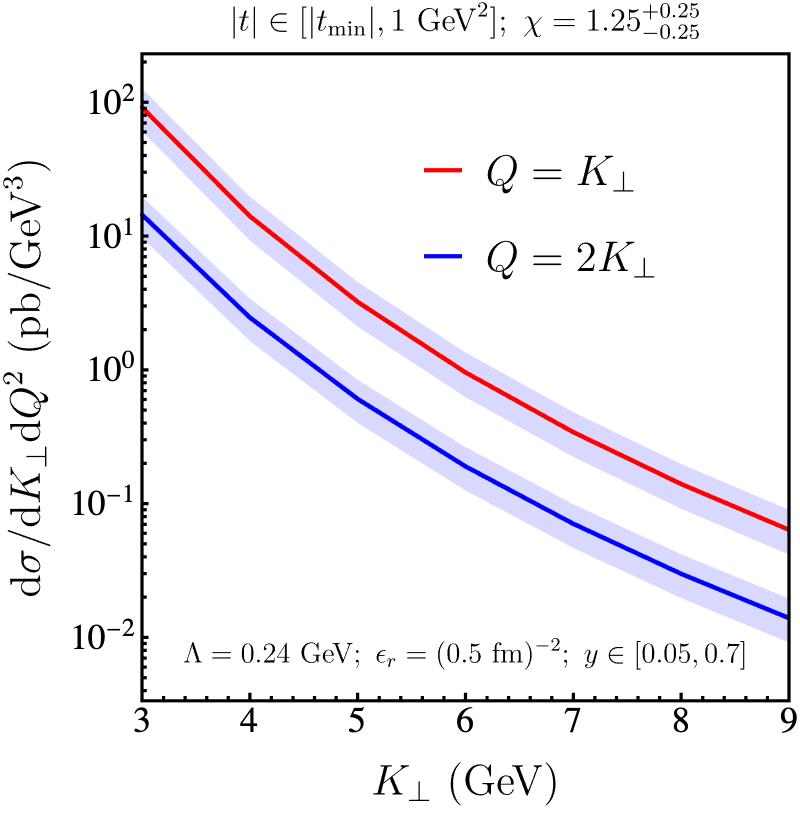}
		\caption{\label{EIC_QK} Predictions for exclusive diffractive dijet production in $ep$ collisions at the EIC for $Q=K_\perp$ and $Q=2 K_\perp$.}
	\end{minipage} \hfill
\end{figure}

The photoproduction result can also be translated into predictions for Ultra-Peripheral Collisions (UPCs) in p-Pb and Pb-Pb collisions at the LHC upon folding in the appropriate photon distribution inside a Pb nucleus, cf.\ \cite{Hagiwara:2017fye}. However, if in such collisions $K_\perp$ values are reached that are much larger than $Q_s$, then the saturation description may no longer be appropriate. Also it is important that the dijet pair has a rapidity gap in order to ensure a diffractive process. The only currently available UPC jet production data is by the ATLAS collaboration \cite{Angerami:2017kot}, where such a rapidity gap condition is not imposed however. The diffractive contribution was estimated to be at most at the 5\% level for the small $x$ part of the data \cite{Guzey:2020ehb} and it may thus not be surprising that our GTMD model cannot describe those ATLAS UPC data, underestimating the cross section by two or three orders of magnitude. In contrast, the collinear factorization description of \cite{Guzey:2018dlm} at NLO is able to describe the ATLAS data well. Predictions for UPCs in a kinematic regime appropriate for our GTMD model will be considered elsewhere.

\section{Conclusions}
We have considered a small-$x$ model for gluon GTMDs that is based on the MV model with an impact parameter dependent saturation scale, introducing a few free parameters in order to obtain a good fit to the H1 data on diffractive dijet production in electron-proton collisions. The values of the free parameters turn out to be reasonable from the physical interpretation point of view. With this we obtain a good description of the $|t|$ dependence of the data and a reasonable description of the jet transverse momentum dependence. This provides confidence that the gluon GTMD description is appropriate for this process in the examined kinematic range. With this model we provide predictions for the EIC for both electroproduction in a somewhat different kinematic regime and for photoproduction which has a much higher cross section. We hope that this will allow further tests of the underlying GTMD description. 

We also have addressed some theoretical issues known from GPD and small $x$ studies that are relevant for small-$x$ gluon GTMD models with an impact parameter dependent saturation scale. First there is the issue that considering the impact parameter dependence requires the target (and the dipole) to be sufficiently localized in transverse coordinate space. This in turn requires consideration of the dipole frame with large $P^+$ and hence sufficiently large center of mass energy. Second, the dipole size has to be much smaller than the size of the impact parameter profile considered. This requires consideration of the correlation limit for dijet production, in which $\Delta_\perp=k_{1\perp}+k_{2\perp}$ is much smaller than $K_\perp=(k_{1\perp}-k_{2\perp})/2 \approx k_{1\perp} \approx k_{2\perp}$, where the latter scale determines the relevant dipole sizes. On the other hand, $K_\perp$ should not be so large that one is outside the saturation regime. 
Under these kinematic conditions the MV model with impact parameter dependent saturation scale is expected to be an appropriate model for the gluon GTMDs probed. The fairly good description of the H1 data gives support to this GTMD picture and we hope it can be scrutinized further with future data from EIC and LHC. 

\acknowledgments
We thank Markus Diehl for valuable discussions and feedback and Yoshikazu Hagiwara for help in reproducing some of his results. The work of C.S.\ is supported by the Indonesia Endowment Fund for Education (LPDP) via a doctoral scholarship.

\end{document}